# Advancing single-atom catalysts: engineered metal-organic platforms on surfaces

Amogh Kinikar,[1],§ Xiushang Xu,[2],§ Takatsugu Onishi,[2] Andres Ortega-Guerrero,[1] Roland Widmer,[1] Nicola Zema,[3] Conor Hogan,[3] Luca Camilli,[4] Luca Persichetti,[4] Carlo A. Pignedoli,[1] Roman Fasel,[1,5] Akimitsu Narita,[2]* Marco Di Giovannantonio[3]*

[1]*Empa, Swiss Federal Laboratories for Materials Science and Technology, nanotech@surfaces Laboratory, 8600 Dübendorf, Switzerland*
[2]*Organic and Carbon Nanomaterials Unit, Okinawa Institute of Science and Technology Graduate University, 904-0495 Okinawa, Japan*
[3]*CNR – Istituto di Struttura della Materia (CNR-ISM), 00133 Roma, Italy*
[4]*Dipartimento di Fisica, Università di Roma "Tor Vergata", 00133 Roma, Italy*
[5]*Department of Chemistry, Biochemistry and Pharmaceutical Sciences, University of Bern, Bern 3012, Switzerland*
§*These authors contributed equally to this work*
*\*Corresponding authors: akimitsu.narita@oist.jp (A.N.), marco.digiovannantonio@ism.cnr.it (M.D.G.)*

## Abstract

Recent advances in nanomaterials have pushed the boundaries of nanoscale fabrication to the limit of single atoms (SAs), particularly in heterogeneous catalysis. Single atom catalysts (SACs), comprising minute amounts of transition metals dispersed on inert substrates, have emerged as prominent materials in this domain. However, overcoming the tendency of these SAs to cluster beyond cryogenic temperatures and precisely arranging them on surfaces pose significant challenges. Employing organic templates for orchestrating and modulating the activity of single atoms holds promise. Here, we introduce a novel single atom platform (SAP) wherein atoms are firmly anchored to specific coordination sites distributed along carbon-based polymers, synthesized via on-surface synthesis (OSS). These SAPs exhibit atomic-level structural precision and stability, even at elevated temperatures. The asymmetry in the electronic states at the active sites anticipates the enhanced reactivity of these precisely defined reactive centers. Upon exposure to CO and $CO_2$ gases at low temperatures, the SAP demonstrates excellent trapping capabilities. Fine-tuning the structure and properties of the coordination sites offers unparalleled flexibility in tailoring functionalities, thus opening avenues for previously untapped potential in catalytic applications.

## Introduction

Catalysts play a critical role in modern society: from pharmaceuticals to metallurgy, improving the catalysts involved has led to better products that are more sustainably manufactured. Many catalysts are transition or rare earth metals[1], and one way to improve their efficiency, while simultaneously making them more sustainable, is to use their smallest functional units leading to single atom catalysts (SACs)[2]. This maximizes the utilization of the function-bearing entities yielding an improved atom economy[3,4]. In SACs, the atoms are dispersed on and are supported by surfaces of host materials (Fig. 1a), typically ceramics such as zeolites[5], oxides[6], 2D materials such as graphene[7–9], other metals such as copper[10–13], as well as metal-organic frameworks[14–16]. The isolated single atoms exhibit properties distinct from conventional metal clusters and nanoparticles, and closely resemble the reactive centres in solution or gas phase[17]. As heterogeneous catalysis reaches this atomic frontier, the exact chemical environment of the single atoms defines their catalytic properties[18–20]. This leads to several challenges[2]: attaining a high density of individual atoms without the formation of metal clusters, ensuring that the majority of these atoms are in a desirable chemical state on the surface, and developing adaptable platforms compatible with various metals. Consequently, achieving a high-density uniformly dispersed arrangement of single atoms, each with an atomically precise local environment, remains a longstanding objective. This goal also parallels the function of biological enzymes, where individual metal atoms are precisely coordinated within a protein structure, enabling a broad spectrum of catalytic actions[21,22].

Here, we introduce a surface-supported one-dimensional organic polymer with terpyridine (tpy) functional motifs[23] as side groups. These motifs can be chelated, post-synthesis, with a variety of metal atoms, making these polymers versatile atomically precise SACs (Fig. 1b). The atomic precision is achieved by triggering selective surface-catalysed reactions of molecular precursors on single crystal Au(111) surfaces, a process often called on-surface synthesis[24]. These organic polymers have several structural advantages due to the achieved structural precision, for instance, by having all the metal atoms in the same chemical environment, the elucidation of the structure-property relationship is simplified[25]. Compared to coordination polymers[26,27], which are also atomically precise, these polymers have the metal coordinated to the side functional groups, thus making metal coordination independent from polymerization. Not only does this enable the post-synthesis metal chelation, but it allows for asymmetrically coordinated metal atoms. Breaking the symmetry of the metal centres can be critical in improving the catalytic efficiency[21,28–30]. Additionally, the $N_3$ coordination achieved using the terpyridine motif also provides ample steric access to the coordinated atom. Lastly, although future catalysts for actual applications may rely on nanoparticle-based materials, having the SACs on a clean, crystalline and atomically flat substrate such as Au(111) facilitates their fundamental characterization using powerful tools of surface science[31], particularly bond resolved scanning probe imaging that allows the direct visualization of active sites and bonding motifs.

We demonstrate the atomically precise SAC using a 4'-{4-(2,7-dibromoanthracen-9-yl)phenyl}-2,2':6',2''-terpyridine (DBAP-tpy) precursor (Fig. 1c) that undergoes dehalogenative aryl-aryl coupling on an Au(111) surface in ultrahigh vacuum (UHV) conditions. Post-synthesis, these polymers are coordinated with cobalt atoms. The resulting material is characterized by scanning tunneling microscopy (STM, example large-scale image in Fig. 1d) and non-contact atomic force microscopy (nc-AFM) with functionalized tips, complemented with density functional theory (DFT) simulations (performed with AiiDAlab based applications)[32]. This high-resolution investigation unveils notable changes within the active sites during the synthetic steps and after exposure to gases such as CO and $CO_2$ (Fig. 1e), providing fundamental insights into the potential of this novel material as a SAC with asymmetric reactive centres (Fig. 1f).

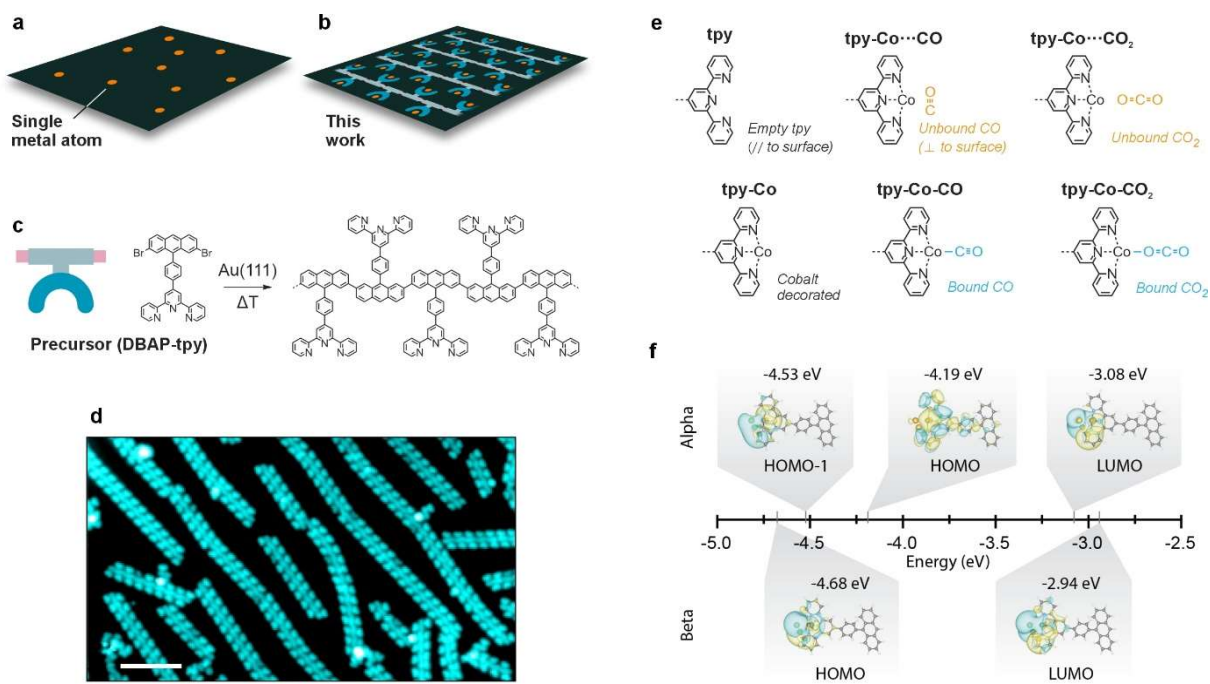

**Fig. 1 | Engineering single atom catalysts. a**, Sketch of surface-adsorbed single atoms (orange dots). **b**, Illustration of our SAC where the organic part is represented by 1D polymers adsorbed on a surface, and the single metal atoms decorate the coordination sites (U-shaped motifs). **c**, The use of a DBAP-tpy molecular precursor enables the selective growth of 1D polymers via on-surface synthesis. **d**, Large-scale experimental STM image of the SAC attained in this work. Scale bar: 10 nm. **e,** Terpyridine functionalization motifs that are discussed in the text. **f,** Gas phase DFT molecular orbital diagram of the frontier orbitals for a molecular unit with an $Au_4$ cluster in proximity of the tpy-Co(II) (geometry in Supplementary Fig. 4). The molecule is in a doublet spin state and both alpha and beta spin states are reported, highlighting the broken symmetry of the states in the active site.

**On-surface fabrication of the atomically precise SAC**

The atomically precise SAC was fabricated using a three-step process. First, the DBAP-tpy precursor molecules were deposited on a clean Au(111) surface held at 200 °C under UHV conditions. This temperature promotes the debrominative homo-coupling yielding 1D polymers (Fig. 1c)[33]. High-resolution imaging of one of these polymer chains reveals a polyanthracene backbone with phenylene-terpyridine side moieties (Fig. 2a,b), as expected. Due to the precursor design and surface confinement, the steric hindrance operated by the tpy groups guarantees high selectivity in the covalent coupling and only linear chains are formed[34,35]. While the tpy units appear as a Y-shaped feature in STM, nc-AFM with a CO-functionalized tip reveals their intramolecular structure as consisting of three planar rings. The central pyridine ring is slightly pulled up by the neighboring phenylene, which is rotated compared to the surface plane due to the hydrogen repulsion with the adjacent anthracene. Such rotation induces a periodic height modulation in the polyanthracene backbone, as highlighted in Fig. 2b. The nc-AFM image of a tpy unit (Fig. 2c) matches perfectly with the simulated one (Fig. 2d) obtained from the DFT-optimized geometry of the structure in Fig. 2e, indicating that the peripheral nitrogen atoms are pointing towards the polymer backbone. We also notice that the tpy units are slightly tilted towards the polymer long axis, probably due to a better match with the underlying surface lattice achieved in this configuration.

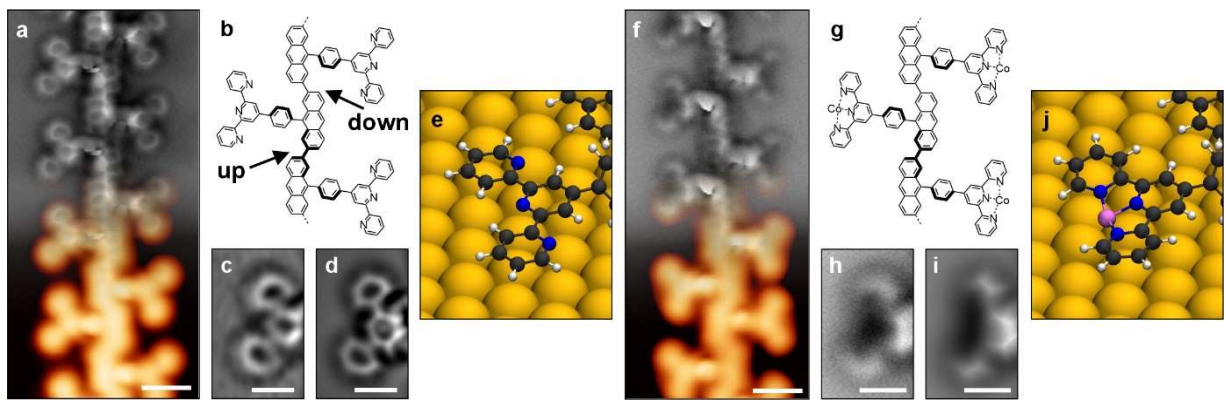

**Fig. 2 | Atomically precise SAC. a,f,** Experimental STM (orange color scale, $I_t$ = 100 pA, $V_b$ = –0.3 V (**a**), $I_t$ = 50 pA, $V_b$ = –0.1 V (**f**)) and nc-AFM (grey color scale, $\Delta z$ = +170 pm (**a**), $\Delta z$ = +190 pm (**f**)) images of the 1D polymer obtained from the DBAP-tpy precursor on Au(111) via the OSS approach, before (**a**) and after (**f**) decoration with cobalt atoms. **b,g,** Chemical schemes of the obtained polymers. Bold segments indicate higher parts (i.e., further from the surface plane). **c,h,** Zoom-in experimental nc-AFM images ($\Delta z$ = +150 pm (**c**), $\Delta z$ = +200 pm (**h**)) of a tpy (**c**) and a tpy-Co (**h**) unit. **d,i,** Simulated nc-AFM images of a tpy (**d**) and a tpy-Co (**i**) unit obtained from the structures in (**e**) and (**j**), respectively. **e,j** Zoom-in of the DFT-optimized geometries of a polymer segment on Au(111), without (**e**) and with (**j**) cobalt atoms coordinated to the tpy sites. Scale bars: 1 nm (**a,f**), 0.5 nm (**c,d,h,i**).

A sample with medium/high polymer coverage features chemisorbed bromine atoms that coexist with the polymer chains and stabilize them in islands (Supplementary Fig. 1a). If cobalt atoms are deposited on the surface at this stage, a cobalt-bromine complex coordinates to the terpyridine (tpy-Co-Br, Supplementary Fig. 1b and S2). To achieve the targeted active sites featuring only cobalt as coordinated atom, the second step of SAP preparation consists in the removal of bromine atoms. If this is attempted by annealing the surface at higher temperatures, some intramolecular transformations occur in the polymer before complete bromine desorption. Hence, we used an alternative, well established procedure, which exploits atomic hydrogen[36]. This treatment (see Methods for the experimental details) successfully removes all bromine atoms from the surface and leaves the polymer chains (nearly) unaltered (Supplementary Fig. 1c and 3).

The third procedural step involves the controlled dosing of cobalt atoms, executed with the sample held at 150 °C to enhance surface diffusion and prevent cobalt clustering into islands. Optimal amount of cobalt ensures the saturation of all tpy units without excess. We meticulously calibrated the cobalt surface coverage to achieve a precise 1:1 ratio between tpy and Co. Subsequently, the sample underwent annealing at 200 °C to enhance order and cleanliness post the aforementioned steps. As a result, the Y-shaped features observed by STM appear now less dark and less sharp in their center (Fig. 2f) as compared to the pristine polymers (Fig. 2a). nc-AFM reveals that the tpy moieties are no longer planar, as their constituent rings appear tilted down at the center of the unit (Fig. 2f,h). This is consistent with the presence of a cobalt atom coordinated to the tpy and closer in height to the surface plane, as evidenced by the good match with the simulated nc-AFM image (Fig. 2i) of a tpy-Co unit (Fig. 2j) and in agreement with a previous study using iron-coordinated tpy on a silver surface[37].

We affirm the successful fabrication of the targeted SAC, with cobalt loading that amounts to 10.9 wt%, or 1.7 at%. Samples with medium molecular coverage facilitated the growth of the SAP with quasi-evenly spaced chains adsorbed predominantly on the fcc regions of the Au(111) 22×√3 (herringbone) reconstruction (Supplementary Fig. 1d). In such samples, the average density of active sites (tpy-Co) is approximately 0.15 per nm², corresponding to one every 6.6 nm². The DFT-calculated binding energy of cobalt atoms within the tpy pockets is 3.8 eV, obtained as energy difference between an optimized geometry with coordinated cobalt (e.g Fig. 2j) and another where the cobalt atom is on the substrate, far from the coordination site. Given that the sample underwent a temperature of 200 °C during preparation, this substantial binding energy anticipates the high stability of the active sites. The well-defined entities achieved with our protocol feature cobalt atoms in an asymmetric coordination environment. To visualize such broken symmetry scenario, we have computed the spatial distribution of the frontier molecular orbitals via DFT single-point energy simulations. In the doublet spin configuration (more stable than the quartet, see Supplementary Table 1), frontier states exhibit significant hybridization with the orbitals from the gold substrate (represented by a $Au_4$ cluster in these calculations) (Fig. 1f). Without the gold, the orbitals of the tpy-Co complex in gas phase show hybridization of π and d orbitals from the transition metal near the frontier molecular orbitals (Supplementary Fig. 5 and 6), unlike the pure π-π* frontier orbitals of a Co(II) porphyrin (Supplementary Fig. 7). When comparing tpy-Co and Co(II) porphyrin, both molecules have an active site at the Co atom. However, the tpy-Co coordination environment is unsaturated and results in a broken symmetry of the orbitals, providing an optimal spatial distribution for catalytic activity[38]. This is reflected in their electronic properties in the gas phase. In tpy-Co, d and π orbitals are hybridized near the Fermi level, whereas in porphyrin, the d orbitals are not close to the frontier orbitals, being situated both below and above the Fermi level. The asymmetric orbital arrangement obtained in our system can facilitate more efficient electron transfer and interaction with reactants, potentially leading to an improved catalytic activity.

**Exposure to CO**

Cobalt, a pivotal metal in industrial processes and catalysis, plays a crucial role in diverse reactions, such as Fischer-Tropsch synthesis[39], oxidation of methane[40], steam reforming of ethanol[41] and methane[42], and is prohibitively expensive. Our innovative SAC comprises a minimal amount of this valuable material – yet our protocol is also applicable to other precious metals – and presents an ideal opportunity to explore the performance of low cobalt doses under gas exposure. Our attention is directed towards carbon monoxide (CO) and carbon dioxide ($CO_2$) as model systems due to their central role in the mentioned industrial processes and $CO_2$ reduction.

We exposed our SAC to CO at a partial pressure of $3\times10^{-8}$ mbar for 90 s, with the Au(111) sample maintained at 9.0 K. This low dose typically results in a very low surface density of physisorbed CO molecules, used to functionalize the STM tip for improved imaging. The resulting STM image (Fig. 3a) reveals variations in tpy-Co sites, some appearing empty, while others exhibit dim or bright dots. Concurrent nc-AFM imaging clarifies these features (Fig. 3b). Empty units correspond to the tpy-Co sites described above. Dim dots in STM align with bright circular protrusions in nc-AFM (yellow arrows in Fig. 3b), resembling CO molecules physisorbed on the Au(111) surface and trapped between the polymer chains (white arrows in Fig. 3b). Hence, these units feature CO molecules standing perpendicular to the surface and stabilized by the cobalt atom without direct bonding (tpy-Co⋯CO in Fig. 1e). Remarkably, less than 1% of units exhibit a bright center in STM (green circle in Fig. 3a). The corresponding nc-AFM image displays two faint features extending from the tpy (green circle in Fig. 3b), representing a CO molecule nearly parallel to the surface and covalently bound to the cobalt atom, i.e., tpy-Co-CO (Fig. 1e and 3c). The simulated nc-AFM image (Fig. 3e) of a DFT-optimized tpy-Co-CO unit (Fig. 3f) perfectly aligns with the experimental image (Fig. 3d), confirming our assignment and consistently with a similar study on isolated molecules containing tpy-Fe moieties on Ag(111)[43].

Significantly, elevating the CO dose ($5\times10^{-8}$ mbar for 15 minutes) followed by a wobblestick annealing (WA, see Methods) for 15 minutes results in a sample, where 89% of the active sites exhibit bound CO molecules, specifically in the form of tpy-Co-CO. This pronounced yield, illustrated in the STM images (Fig. 3g and Supplementary Fig. 1e), underscores the impressive efficacy of our SAC in capturing CO. The distinctive appearance of tpy-Co-CO sites in STM imaging demonstrates the potential for detecting CO at low temperatures, obviating the need for nc-AFM imaging. It's noteworthy that the WA step is crucial for promoting CO diffusion and binding to cobalt, as the tpy-Co-CO sites were only 1% before this treatment, with most CO adsorbed on the bare Au(111) or adjacent to the polymer chains. This highlights a key advantage of our tpy-based SAC – its open structure. Gas molecules are not obliged to directly interact with the active site upon approaching the surface; instead, they can be adsorbed at any location and diffuse until reaching the cobalt atoms. Such a mechanism is absent in closed pores or macrocycles. Complete CO desorption from the Au(111) surface and the tpy-Co sites is achieved after annealing the sample to room temperature, preserving the SAC for cyclic reuse without harm.

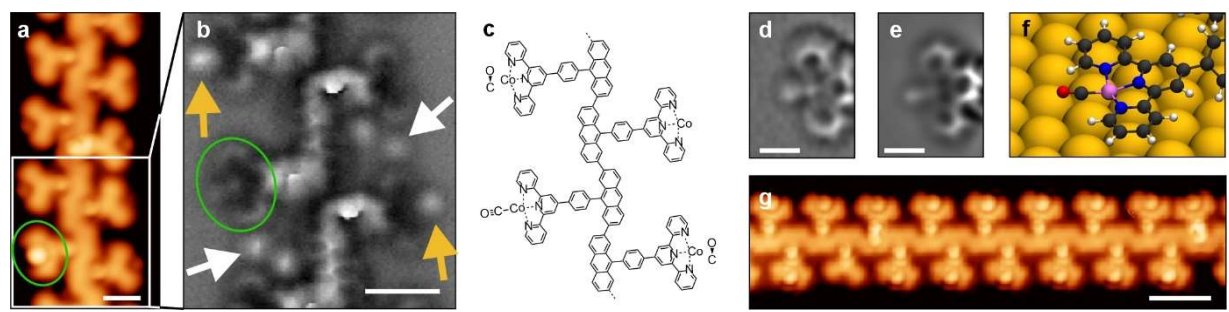

**Fig. 3 | Trapping of CO. a,** Experimental STM image ($I_t$ = 70 pA, $V_b$ = –0.1 V) of a polymer segment where the tpy-Co sites are either empty, with bound CO, or with unbound CO. **b,** nc-AFM image ($\Delta z$ = +220 pm) of the region highlighted by the white box in **a**. The round protrusions in between tpy units (white arrows) are physisorbed CO molecules stabilized next to the polymer backbone, and have similar appearance to the CO molecules sitting next to the tpy-Co sites (tpy-Co⋯CO, yellow arrows). The green circles (**a,b**) highlight a tpy-Co-CO unit, where CO is bound to the active site. **c,** Molecular scheme of the segment and active site occupations in **b**. **d,** Experimental nc-AFM image ($\Delta z$ = +140 pm) of a tpy-Co-CO site. **e,** Simulated nc-AFM image of the tpy-Co-CO unit obtained from the structure in **f**. **f,** Zoom-in of the DFT-optimized geometry of a polymer segment on Au(111), with tpy-Co-CO units. **g,** Experimental STM image ($I_t$ = 100 pA, $V_b$ = –0.1 V) of a polymer segment with nearly all units featuring bound CO molecules. Scale bars: 1 nm (**a,b**), 0.5 nm (**d,e**), 2 nm (**g**).

## Exposure to $CO_2$

When exposing our SAP to $CO_2$ ($5\times10^{-8}$ mbar for 15 minutes) with the sample held at 30 K, the resulting surface exhibited chains with different occupation of the tpy-Co sites. Most of them show large, round features (Fig. 4a), which appear as elongated, rod-like objects in nc-AFM (Fig. 4b,c). Those next to the tpy-Co are slightly nonplanar. The simulated nc-AFM image of a DFT-optimized tpy-Co⋯$CO_2$ unit (Fig. 4e,f) shows an excellent match with the experimental images. The rod-like feature of the $CO_2$ is fully reproduced, including its nonplanar appearance due to a weak interaction with the Co. The calculated distance between cobalt and the nearest oxygen is 2.81 Å. This value, significantly larger than the distance between cobalt and carbon in the case of tpy-Co-CO units (1.74 Å), suggests that the $CO_2$ is not bound to the cobalt, but weakly stabilized next to it (tpy-Co⋯$CO_2$, see Fig. 1e and Supplementary Fig. 8). The coexistence of "empty" tpy-Co sites with those decorated with CO and $CO_2$ (Fig. 4d), offers a direct comparison of the possible functionalization motifs and highlights the strength of our characterization down to the atomic level. Remarkably, we have further investigated the sample after stepwise annealing and observed the diminishing of $CO_2$ content accompanied by an unexpected increase of CO. This finding offers intriguing perspectives for a possible low-temperature catalytic activity of our SAP in $CO_2$ conversion (see corresponding sections in the SI).

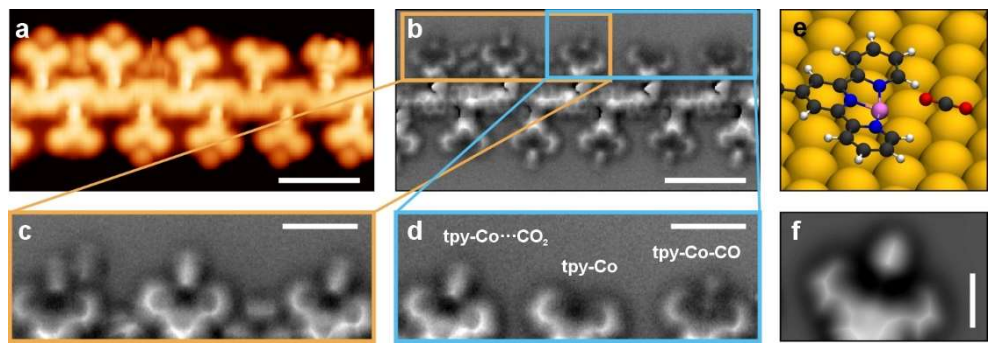

**Fig. 4 | Trapping of CO$_2$. a,b,** Experimental STM ($I_t$ = 100 pA, $V_b$ = –0.1 V) and nc-AFM (Δz = +240 pm) images of a polymer segment where the tpy-Co sites are either empty or with CO$_2$ sitting next to them. A tpy-Co-CO site is also visible (top right). **c,d,** Zoom-in nc-AFM images (Δz = +220 pm) of the areas highlighted by the yellow and cyan boxes in **b**. **e,** DFT-optimized geometry of a polymer segment on Au(111), with tpy-Co⋯CO$_2$. **f,** Simulated nc-AFM image of the tpy-Co⋯CO$_2$ unit obtained from the structure in **e**. Scale bars: 2 nm (**a,b**), 1 nm (**c,d**), 0.5 nm (**e,f**).

## Outlook

We have demonstrated the synthesis of an organic polymer that can stabilize metal atoms through their coordination to its side groups. This leads to an atomically precise SAC. Because our synthesis relies on well-established surface chemistry protocols, it can easily be extended by choosing different molecular precursors. A great strength of on-surface synthesis is the tuneability of the structural and electronic properties of the target products, achieved by changing the molecular precursor used as starting material[24,44–46]. The vast number of well-established on-surface reactions consequently provides a large array of possible atomically precise SAC structures. This allows for the creation of a library of atomically precise SACs greatly aiding the elucidation of the structure-property relationships. Moreover, surface synthesized structures can be transferred from their growth substrate to arbitrary substrates using several different transfer protocols[47–49]. These same protocols are available to the surface synthesized SACs. The terpyridine-bearing polymers can also be chelated in solution after their transfer, further extending their versatility. Finally, the present work was focused on structural characterization of the polymers and the interaction of the metal atoms with gas molecules. This necessitated sub-monolayer coverages of the polymers on the Au(111) single crystal surface. However, large scale synthesis of these polymers is possible using chemical vapour deposition techniques[50] and adopting Au(111) thin films on Mica or Au nanoparticles as substrates, thereby providing a path towards mass production and commercial applications.

## Conclusions

Creating atomically dispersed catalyst sites, with high surface density and precise coordination is a paramount goal within the field of single atom catalysis. Here, we presented a unique strategy, building on recent advances in the field of on-surface synthesis, to create an atomically precise SAC based on organic polymers with terpyridine side motifs that can be chelated, post synthesis. The resulting asymmetric coordination of the metal atoms makes these structures highly promising catalysts. Moreover, using high resolution scanning probe microscopy, we could experimentally demonstrate the exact binding configuration of different compounds to the metal atom and support our experimental finding with ab initio calculations. Our SAC platform exhibits excellent trapping ability at low temperature towards CO and CO$_2$ with approx. 90% and 60% yield, respectively, and fully releases the gases upon annealing at room temperature, allowing for continuous reuse without poisoning. Our work introduces a pathway towards fabrication of a diverse class of atomically precise SACs, the investigation of whose catalytic properties would yield fundamental insights into how catalysts work. In the future we envisage that such atomically precise catalysts could be rationally designed to specifically catalyze chosen reactions, thus approaching the versatility and specificity of biological enzymes.

**Methods**

*General Methods for precursor synthesis*
All reactions working with air- or moisture-sensitive compounds were carried out under argon atmosphere using standard Schlenk line techniques. All starting materials, reagents, and solvents were purchased from commercial sources and used as received unless otherwise noted. 5-Bromo-2-(4-bromobenzyl)benzaldehyde (**3**) was prepared according to a previously reported procedure[51]. Anhydrous dimethylformamide (DMF), tetrahydrofuran (THF), and dichloromethane ($CH_2Cl_2$) were purified by a solvent purification system (GlassContour) prior to use. Thin-layer chromatography (TLC) was done on silica gel coated aluminum sheets with F254 indicator and column chromatography separation was performed with silica gel (particle size 0.063-0.200 mm). Nuclear Magnetic Resonance (NMR) spectra were recorded in $CDCl_3$ using Bruker Avance Neo 400 and 500 MHz NMR spectrometers. Chemical shifts ($\delta$) were expressed in ppm relative to the residual of solvents ($CDCl_3$, $^1$H: 7.26 ppm, $^{13}$C: 76.00 ppm). Coupling constants ($J$) were recorded in Hertz. Abbreviations: s = singlet, d = doublet, t = triplet, m = multiplet. High-resolution mass spectra (HRMS) were recorded on a Bruker ultrafleXtreme spectrometer by matrix-assisted laser desorption/ionization (MALDI) with tetracyanoquinodimethane (TCNQ) as the matrix, or on Thermo Scientific LTQ-Orbitrap Mass Spectrometer by electrospray ionization (ESI).

*On-surface experiments*
The on-surface synthesis experiments were performed under ultrahigh vacuum (UHV) conditions with base pressure below $2\times10^{-10}$ mbar. Au(111) substrates (MaTeck GmbH) were cleaned by repeated cycles of Ar$^+$ sputtering (1 keV) and annealing (460 °C). The precursor molecules were thermally evaporated onto the clean Au(111) surface held at 200 °C from quartz crucible heated at 170 °C (deposition rate of ~ 0.5 Å·min$^{-1}$). Atomic hydrogen was dosed onto the surface held at 150 °C by using a gas cracker (Mantis) operated at 70 W with a partial pressure of $2\times10^{-7}$ mbar of molecular hydrogen (Messer, purity 5.0) for 15 minutes. Cobalt atoms were sublimated onto the surface held at 150 °C from a rod (Alfa Aesar, purity 4.5) heated in a home-built e-beam metal evaporator, and the surface coverage was controlled using the STM and nc-AFM images as feedback. After cobalt deposition, the sample was annealed to 200 °C for 20 minutes. CO, $CO_2$, and Ar gases (Linde, purity 4.7, Pangas, purity 4.5, and Messer, purity 5.0, respectively) were dosed onto the cold sample in the STM stage with a partial pressure of $5\times10^{-8}$ mbar for 15 minutes. The sample temperature reached 9.5 K during the exposure with the shields of the STM cryostat kept open. In some cases, we performed an unusual annealing of the sample, to prevent potential contamination from the STM stage that might occur in the case of standard resistive heating. Such wobblestick annealing (WA) consisted in extracting the cold sample from the STM stage with the (room temperature) pincer tool that is usually operated for transfers (namely, the wobblestick), and holding the sample on its pincer for a specific time. Afterwards, the sample is reinserted into the cold STM stage and cooled down to 4.7 K. While not allowing any temperature readout, this method guarantees the cleanest annealing conditions.

## STM and nc-AFM imaging

STM images were acquired with a low-temperature STM/nc-AFM (Scienta Omicron) operated at 4.7 K in constant-current mode using an etched tungsten tip. The scanning parameters are indicated in the captions, with bias voltages referred to the sample. nc-AFM measurements were performed at 4.7 K with a tungsten tip placed on a qPlus force sensor[52]. The tip was functionalized with a single CO molecule at the tip apex picked up from the previously CO-dosed surface[53]. The sensor was driven at its resonance frequency (27405 Hz) with a constant amplitude of 70 pm. The frequency shift from resonance of the tuning fork was recorded in constant-height mode using Omicron Matrix electronics and HF2Li PLL by Zurich Instruments. The $\Delta z$ reported in the captions is positive (negative) when the tip-surface distance is increased (decreased) with respect to the STM setpoint at which the feedback loop is open ($I_t$ = 100 pA, $V_b$ = –5 mV, with the tip located on an empty area of the Au(111) surface).

## Computational details

### On surface

All calculations were performed with AiiDAlab [32] apps based on the DFT code CP2K[54]. The surface/adsorbate system was modeled within the repeated slab scheme, with a simulation cell containing up to 1500 atoms, of which 4 atomic layers of Au along the [111] direction and a layer of hydrogen atoms to passivate one side of the slab in order to suppress one of the two Au(111) surface states. 40 Å of vacuum was included in the simulation cell to decouple the system from its periodic replicas in the direction perpendicular to the surface. The size of the cell was 38.3×51.1 Å$^2$ corresponding to 260 Au(111) surface unit cells. The electronic states were expanded with a TZV2P Gaussian basis set[55] for C and H species and a DZVP basis set for Au and Co species. A cutoff of 600 Ry was used for the plane wave basis set. Norm-conserving Goedecker-Teter-Hutter pseudopotentials[56] were used to represent the frozen core electrons of the atoms. We used the PBE parameterization for the generalized gradient approximation of the exchange correlation functional[57]. To account for van der Waals interactions, we used the D3 scheme proposed by Grimme[58]. To obtain the equilibrium geometries, we kept the atomic positions of the bottom two layers of the slab fixed to the ideal bulk positions, and all other atoms were relaxed until forces were lower than 0.005 eV/Å. The Probe Particle model[59] was used to simulate nc-AFM images.

### Gas phase

From the optimized structure of tpy-Co(II) on an Au(111) surface discussed in the main text, a representative model containing an Au$_4$ cluster and the tpy-Co(II) molecule was defined (see its geometry in Supplementary Fig. 4). DFT single-point energy simulations were performed using Gaussian 16[60] with the PBE0 exchange-correlation functional[61]. The def2-TZVP basis set[62] was applied to all elements, with pseudopotentials employed for the Au atoms. A visualization of the resulting frontier molecular orbitals is reported in Fig. 1f in the main text.

The spin configuration of the system tpy-Co(II) + Au$_4$ cluster was evaluated for two multiplicities of Co (M = 2 and 4), with the doublet spin state identified as the lowest energy configuration (Supplementary Table 1). Geometry optimization calculations were performed on the tpy-Co(II) molecule in the gas phase, excluding the Au$_4$ cluster, using both spin multiplicities as well (Supplementary Table 2 and Supplementary Fig. 5 and 6). Additionally, molecular orbitals of Co(II) porphyrin were computed for comparison with those of the tpy-Co(II) complex at gas-phase (Supplementary Fig. 7).

**Data availability**
The data supporting the findings of this study are available within the paper and its Supplementary Information.

**Acknowledgements**
We acknowledge the European Union, Next Generation EU, Mission 4, Component 1, CUP B53D23013760006 and the Italian Ministry of University and Research (MUR) for the PRIN 2022 (project No. 2022JW8LHZ, ATYPICAL), the Okinawa Institute of Science and Technology Graduate University, the CNR-JSPS (Italy-Japan) bilateral project 2023-2024 (project No. JPJSBP120234004, ACCESS), the CNR Short Term Mobility program, the Swiss National Science Foundation under Grant No. 212875, the NCCR MARVEL funded by the Swiss National Science Foundation (205602). Computational support from the Swiss Supercomputing Center (CSCS) under project ID s1267 is gratefully acknowledged. We acknowledge PRACE for awarding access to the Fenix Infrastructure resources at CSCS, which are partially funded from the European Union's Horizon 2020 research and innovation program through the ICEI project under the grant agreement No. 800858. We are thankful to Lukas Rotach (Empa) and Massimiliano Rinaldi (CNR-ISM) for their excellent technical support during the experiments.

**Author contributions**
Conceptualization: A.N., M.D.G.; Funding acquisition: C.A.P., R.F., A.N., M.D.G.; Investigation: A.K., X.X., T.O., A.O.G., R.W., C.H., C.A.P.; Project administration: R.F., A.N., M.D.G.; Resources: R.W., N.Z., C.H., C.A.P., R.F., A.N., M.D.G.; Supervision: R.F., A.N., M.D.G.; Visualization: A.K., A.O.G., C.A.P., M.D.G.; Writing – original draft: A.K., C.A.P., R.F., A.N., M.D.G.; Writing – review & editing: all authors.

**Competing interests**
The authors declare that they have no competing interests.

# Supplementary Information

## for

## Advancing single-atom catalysts: engineered metal-organic platforms on surfaces

**Content:**

    Complete synthetic procedure for DBAP-tpy
    Additional experimental and computational results
    $CO_2$ conversion into CO
    Control experiments
    NMR spectra
    Supplementary references

**Complete synthetic procedure for DBAP-tpy**

DBAP-tpy (**1**) was synthesized as shown in Scheme S1. Diiodobenzene **2** was lithiated and reacted with aldehyde **3**. Subsequently, the obtained hydroxy intermediate was treated with BF$_3$·OEt$_2$ and oxidized by 2,3-dichloro-5,6-dicyano-1,4-benzoquinone (DDQ) to afford 2,7-dibromo-9-(4-iodophenyl)anthracene (**4**) in 46% yield. Then, **4** was subjected to regioselective halogen-metal exchange reaction with isopropylmagnesium chloride at –78 °C, and reacted with dimethylformamide (DMF) to produce aldehyde **5** in 68% yield. Finally, DBAP-tpy **1** was obtained from **5** and 2-acetylpyridine according to a previously reported procedure[S1] in 45% yield.

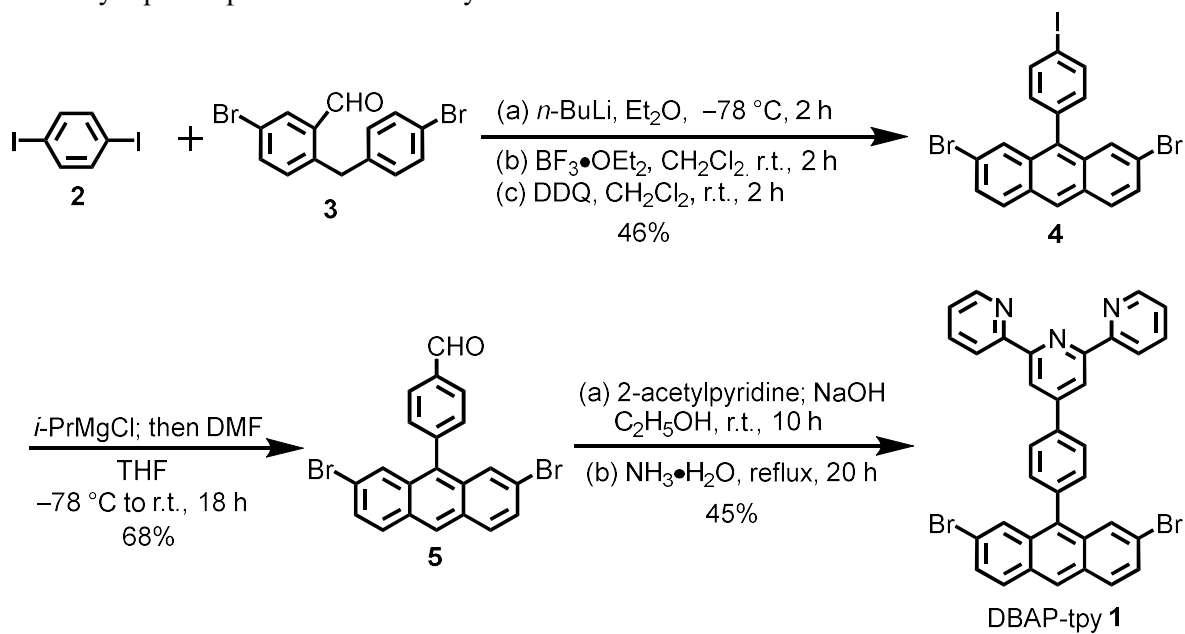

**Scheme S1.** Synthesis of DBAP-tpy **1**.

2,7-dibromo-9-(4-iodophenyl)anthracene (**4**)

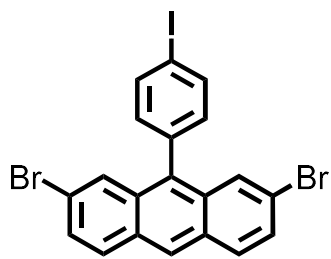

To a solution of 1,4-diiodobenzene (**2**) (100 mg, 0.303 mmol) in anhydrous tetrahydrofuran (THF) (10.0 mL), *n*-butyllithium (1.6 M in hexane, 0.20 mL, 0.32 mmol) was added dropwise at –78 °C under argon atmosphere. After reaction mixture was stirred for 2 h, 5-bromo-2-(4-bromobenzyl)benzaldehyde (**3**) (106 mg, 0.303 mmol) was added. Then, the resulting mixture was stirred overnight at room temperature. The reaction was quenched with saturated aqueous solution of NH$_4$Cl (10 mL) and extracted with CH$_2$Cl$_2$ (10 mL) for three times. The organic phases were combined, washed with brine, dried with MgSO$_4$, and evaporated. The residue was subsequently dissolved in dry CH$_2$Cl$_2$ (10 mL), and BF$_3$·OEt$_2$ (0.05 mL, 0.4 mmol) was added at 0 °C under argon atmosphere. Then, the reaction mixture was stirred 2 h at 0 °C. The resulting mixture was quenched with saturated aqueous solution of NaHCO$_3$ (10 mL) and extracted with CH$_2$Cl$_2$ (10 mL) for three times. The organic phases were combined, washed with brine, dried with MgSO$_4$, and evaporated. The residue was then dissolved in CH$_2$Cl$_2$ (30 mL) and DDQ (209 mg, 0.909 mmol) was added to the solution, followed by stirring at room temperature for 2 h. The reaction mixture was poured into water (10 mL) and extracted with CH$_2$Cl$_2$ (10 mL) for three times.

Then, organic phases were combined, washed with brine, dried over MgSO$_4$, and evaporated. The residue was purified by silica gel column chromatography (eluent: hexane) to give the title compound (70 mg, 46% yield) as light yellow solid. $^1$H NMR (500 MHz, CDCl$_3$) $\delta$ 8.43 (s, 1H), 7.95 (d, $J$ = 7.9 Hz, 2H), 7.90 (d, $J$ = 9.0 Hz, 2H), 7.75 (d, $J$ = 1.8 Hz, 2H), 7.54 (dd, $J$ = 9.0, 1.9 Hz, 2H), 7.13 (d, $J$ = 7.9 Hz, 2H). $^{13}$C NMR (126 MHz, CDCl$_3$) $\delta$ 137.00, 135.84, 132.95, 131.94, 130.24, 129.15, 128.55, 128.23, 127.21, 126.48, 119.95. HRMS (MALDI-TOF, Positive): $m/z$ Calcd. For C$_{20}$H$_{11}$Br$_2$I$^+$: 535.8672 [M]$^+$, found: 535.8696.

4-(2,7-dibromoanthracen-9-yl)benzaldehyde (**5**)

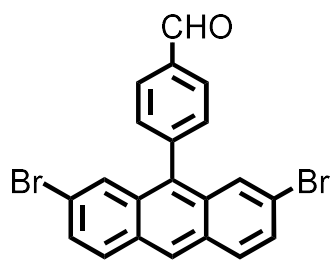

To a solution of compound **4** (570 mg, 1.06 mmol) in anhydrous THF (30 mL) was added *i*-PrMgCl (2.0 M in THF, 0.5 mL, 1.0 mmol) dropwise at –78 °C under argon atmosphere. After stirring the reaction mixture at –78 °C for 2 h, dimethylformamide (DMF) (155 mg, 2.12 mmol) was added. The reaction mixture was allowed to warm to room temperature and stirred overnight at room temperature for 18 h. The reaction was quenched by saturated aqueous solution of NaHCO$_3$ (20 mL) and extracted with CH$_2$Cl$_2$ (30 mL) for three times. The organic layers were combined, washed with brine, dried with MgSO$_4$, and evaporated. The residue was purified by silica gel column chromatography (eluent: hexane: ethyl acetate = 20:1) to afford the title compound (320 mg, 68%) as yellow solid. $^1$H NMR (400 MHz, CDCl$_3$) $\delta$ 10.22 (s, 1H), 8.65 – 8.32 (m, 1H), 8.15 (t, $J$ = 6.0 Hz, 2H), 7.99 – 7.86 (m, 2H), 7.76 – 7.63 (m, 2H), 7.65 – 7.45 (m, 4H). $^{13}$C NMR (101 MHz, CDCl$_3$) $\delta$ 191.90, 143.99, 140.72, 136.49, 136.10, 133.85, 131.94, 131.47, 131.04, 130.16, 129.55, 129.34, 127.99, 127.87, 126.73, 125.35, 122.24, 121.20. MS (MALDI-TOF, Positive): $m/z$ Calcd. For C$_{20}$H$_{12}$Br$_2$O$^+$: 434.92 [M]$^+$, found: 434.92.

4'-{4-(2,7-dibromoanthracen-9-yl)phenyl}-2,2':6',2''-terpyridine (DBAP-tpy **1**)

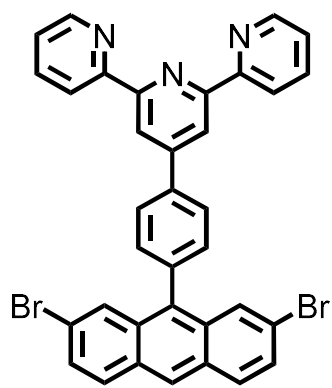

To a solution of NaOH (106 mg, 2.63 mmol) in 10 mL of C$_2$H$_5$OH, compound **5** (210 mg, 0.480 mmol) and 2-acetylpyridine (128 mg, 1.05 mmol) were added. After stirring at room temperature for 10 h, aqueous NH$_3$·H$_2$O (3.0 mL) was added and the mixture was refluxed for 20 h. After cooling to room temperature, ethanol was removed in vacuo. The aqueous phase was extracted with CH$_2$Cl$_2$ (30 mL) for three times. The organic layers were combined, washed with brine, dried with MgSO$_4$, and evaporated. The residue was purified by silica gel column chromatography (eluent: CH$_2$Cl$_2$: methanol = 200:1) to afford the title compound (138 mg, 45%) as light yellow solid. $^1$H NMR (400 MHz, CDCl$_3$) $\delta$ 8.92 (s,

2H), 8.77 (dd, $J$ = 4.8, 1.9 Hz, 2H), 8.73 (dt, $J$ = 8.0, 2H), 8.45 (s, 1H), 8.19 – 8.08 (m, 2H), 7.94 – 7.89 (m, 4H), 7.87 – 7.81 (m, 2H), 7.55 (dd, $J$ = 8.7, 1.6 Hz, 4H), 7.40 – 7.37 (m, 2H), 7.26 (s, 2H). $^{13}$C NMR (101 MHz, CDCl$_3$) $\delta$ 156.08, 149.91, 149.20, 138.49, 138.00, 136.89, 134.92, 131.68, 131.41, 130.12, 129.60, 129.20, 128.49, 127.81, 127.28, 123.90, 121.36, 120.86, 119.08. HRMS (ESI, Positive): $m/z$ Calcd. For C$_{35}$H$_{22}$Br$_2$N$_3$: 642.0143 [M+H]$^+$, found: 642.0143.

**Additional experimental and computational results**

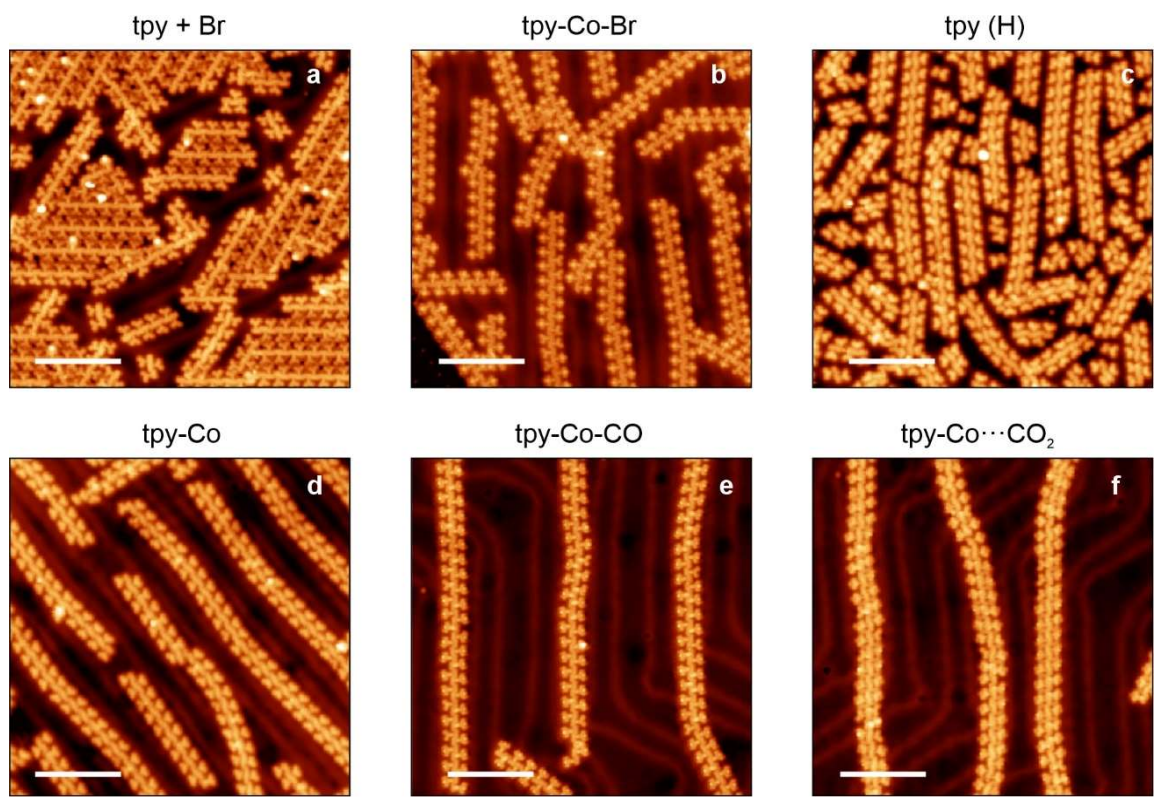

**Supplementary Fig. 1 | Large scale STM images.** STM images of the samples after different preparation processes. **a,** After the growth of polymer chains. Bromine atoms detached from the precursor molecules during the dehalogenative aryl-aryl coupling chemisorb on the surface and are visible in between the polymers. **b,** After dosing cobalt on a sample grown as in **a**. Tpy-Co-Br complexes are formed. **c,** After dosing atomic hydrogen on a sample grown as in **a**. Bromine is no longer visible on the surface. The polymer chains are not packed in islands anymore, but appear well separated from each other. **d,** After dosing cobalt on a sample grown as in **c**. Although not evident from large scale STM images, all tpy sites feature coordinated cobalt atoms (tpy-Co). **e,** After dosing CO on a sample grown as in **d**, and subsequently annealed on the wobblestick for 15 minutes. Nearly all units consist of CO bound to the active site (tpy-Co-CO). **f,** After dosing CO$_2$ on a sample grown as in **d**, and subsequently annealed on the wobblestick for 2 minutes. Most of the units feature CO$_2$ molecules sitting next to the active site (tpy-Co$\cdots$CO$_2$). Scanning parameters: $I_t$ = 70 pA, $V_b$ = –0.50 V (**a**); $I_t$ = 50 pA, $V_b$ = –0.20 V (**b**); $I_t$ = 70 pA, $V_b$ = –0.50 V (**c**); $I_t$ = 200 pA, $V_b$ = –0.02 V (**d**); $I_t$ = 70 pA, $V_b$ = –0.15 V (**e**); $I_t$ = 100 pA, $V_b$ = –0.10 V (**f**). Scale bars: 10 nm (**a-f**).

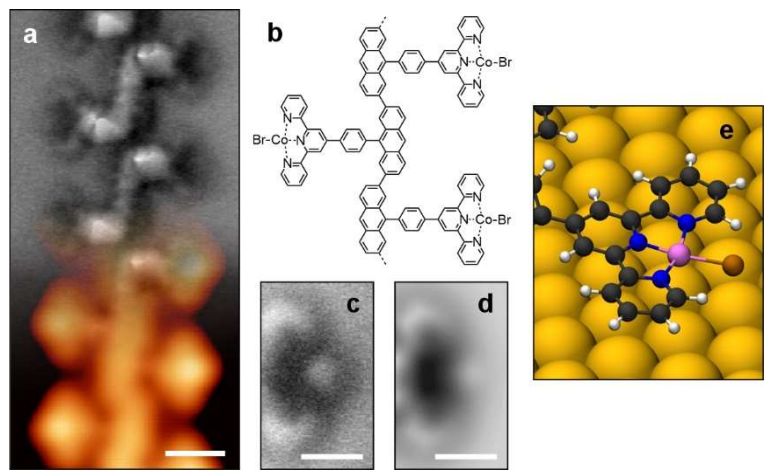

**Supplementary Fig. 2 | Tpy-Co-Br units. a,** Experimental STM ($I_t$ = 100 pA, $V_b$ = –0.1 V) and nc-AFM ($\Delta z$ = +210 pm) images of a polymer segment where the tpy-Co sites are passivated by bromine atoms (tpy-Co-Br). **b,** Chemical scheme of the obtained polymer. **c,** Zoom-in experimental nc-AFM image ($\Delta z$ = +200 pm) of a tpy-Co-Br unit. **d,** Simulated nc-AFM image of tpy-Co-Br unit obtained from the structure in **e**. **e,** Zoom-in of the DFT-optimized geometry of a polymer segment with tpy-Co-Br units on Au(111). Scale bars: 1 nm (**a**), 0.5 nm (**c,d**).

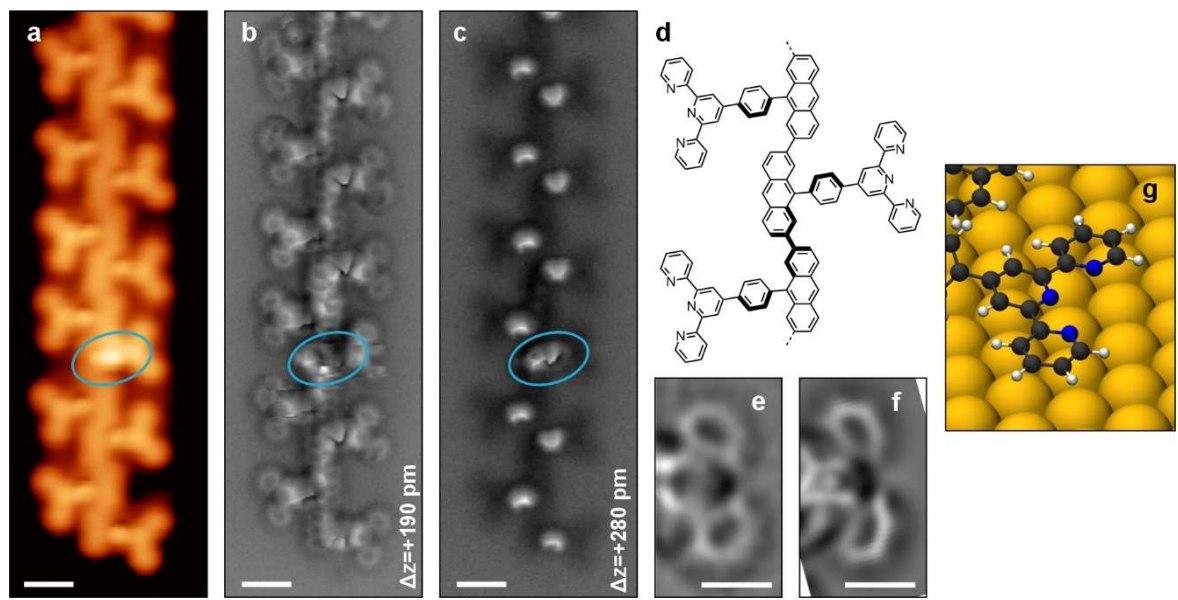

**Supplementary Fig. 3 | Polymer chains after exposure to atomic hydrogen. a-c,** Experimental STM (**a**, $I_t$ = 50 pA, $V_b$ = –0.5 V) and nc-AFM (**b,c**) images of a polymer segment after exposure to atomic hydrogen. Some parts of the polymer backbone present defects (cyan ellipse in **a-c**). These defects appear as bright protrusions in the STM and nc-AFM images and their unambiguous identification is not trivial. We speculate that they can be due to the hydrogenation of some parts of the chain, which however does not affect the structure of the tpy units. **d,** Chemical scheme of the obtained polymer. **e,** Zoom-in experimental nc-AFM image ($\Delta z$ = +220 pm) of a tpy unit. **f,** Simulated nc-AFM image of tpy unit obtained from the structure in **g**. **g,** Zoom-in of the DFT-optimized geometry of a polymer segment on Au(111) with tpy units where the nitrogen atoms are all pointing towards the center of the tpy. Scale bars: 1 nm (**a-c**), 0.5 nm (**e,f**).

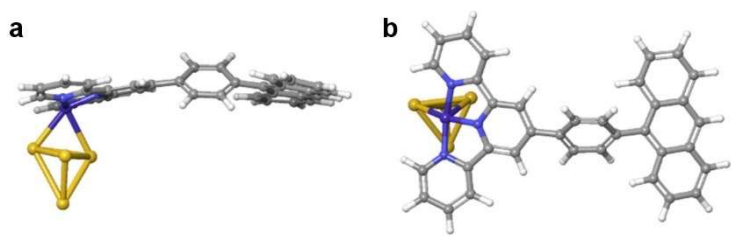

**Supplementary Fig. 4 | Molecular representation of the tpy-Co(II) with a Au$_4$ cluster**. The molecular structure and the gold cluster are extracted from the DFT-optimized geometry of the tpy-Co(II) on a Au(111) surface, and the atomic positions are constrained. Side view (**a**) and top view (**b**) are reported.

**Supplementary Table 1 | DFT Calculated Energy of tpy-Co(II) with the Au$_4$ Cluster Model.**

| System | Energy[eV] |
|---|---|
| Tpy-Co(II) ($s=\frac{1}{2}$)+ Au$_4$ cluster | -93502.972 |
| Tpy-Co(II) ($s=\frac{4}{2}$)+ Au$_4$ cluster | -93502.663 |
| ΔE [$s=\frac{1}{2} - s=\frac{4}{2}$] | -0.309 |

The table presents the DFT calculated single point energies for the tpy-Co(II) complex with the Au$_4$ cluster model. These geometries were derived from a representative model built using periodic calculations (tpy-Co(II) on the Au(111) surface). In this model, both possible spin configurations for Co(II) were considered. Among these, the doublet spin configuration was found to be the most energetically favorable.

**Supplementary Table 2 | DFT Calculated Energies of tpy-Co(II) in Doublet and Quartet Spin Configurations.**

| System | Energy[eV] |
|---|---|
| Tpy-Co(II) ($s=\frac{1}{2}$) | -78727.168 |
| Tpy-Co(II) ($s=\frac{4}{2}$) | -78727.726 |
| ΔE [$s=\frac{1}{2} - s=\frac{4}{2}$] | 0.558 |

The table presents the DFT calculated energies for the tpy-Co(II) complex in both doublet and quartet spin states in the gas phase. The structural coordinates for both spin states were optimized using the PBE0 exchange-correlation functional. The results indicate that in the gas phase, the quartet spin configuration is energetically more favorable than the doublet spin configuration.

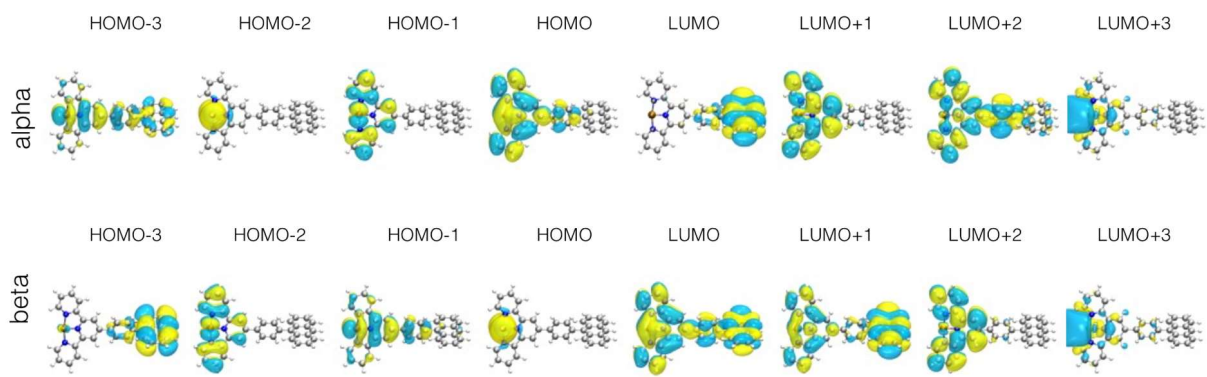

**Supplementary Fig. 5 | Simulated Orbitals of tpy-Co(II) in a doublet spin configuration in gas phase.** The orbitals of tpy-Co(II) complex are obtained from a geometry optimization calculation in gas-phase, where the multiplicity is set as a doublet.

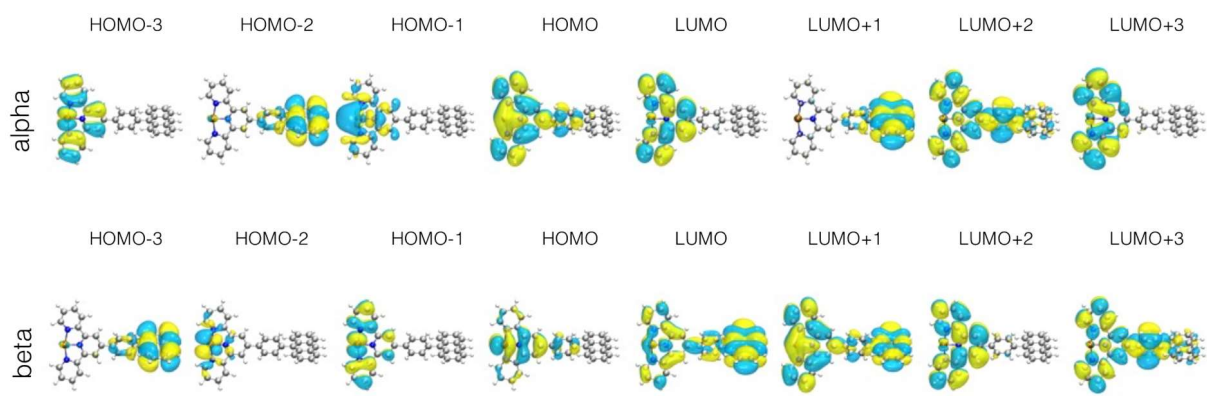

**Supplementary Fig. 6 | Simulated Orbitals of tpy-Co(II) in a quartet spin configuration in gas phase.** The orbitals of tpy-Co(II) complex are obtained from a geometry optimization calculation in gas-phase, where the multiplicity is set as a quartet.

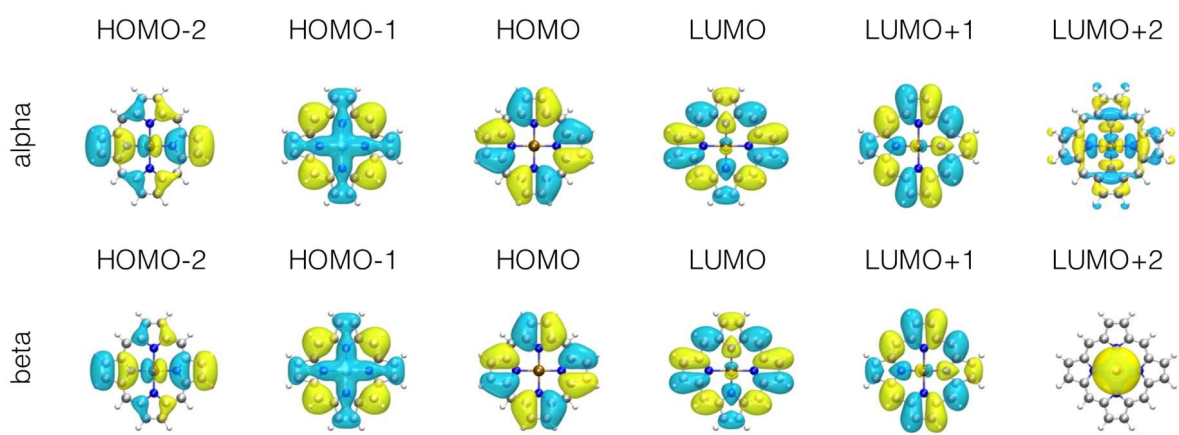

**Supplementary Fig. 7 | Simulated orbitals of Co(II) porphyrin.** The diagram illustrates the frontier molecular orbitals of a Co(II) porphyrin complex, computed using the PBE0 functional with the def2-TZVP basis set. The displayed orbitals demonstrate the characteristic features of the Gouterman model, highlighting the four key π and π* orbitals. Notably, the diagram shows that the unoccupied orbitals include a d$z^2$ orbital that is higher in energy relative to the π* orbitals.

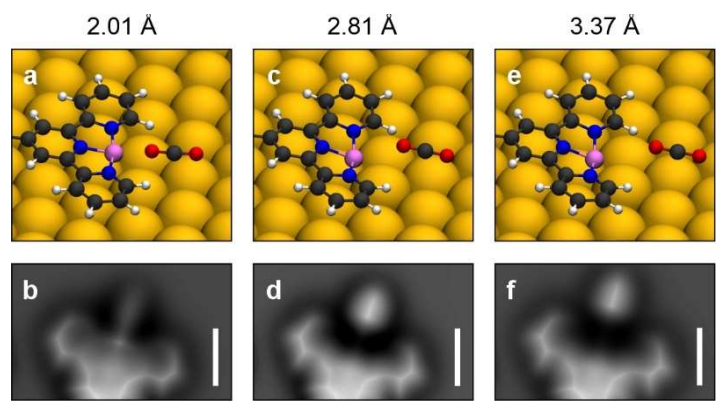

**Supplementary Fig. 8 | Identification of the tpy-Co···CO$_2$ phase. a,c,e,** DFT-optimized geometries with the CO$_2$ molecule at varying distances from the cobalt atom (values on top of the panels indicate the distance between cobalt and the nearest oxygen). The three geometries are obtained after relaxation of different initial guesses. **b,d,f,** Simulated nc-AFM images obtained from the geometries in panels **a**, **c**, and **e**, respectively. The best match with the experimental nc-AFM images in Fig. 4 in the main text is found for an intermediate distance between cobalt and the CO$_2$ of 2.81 Å. Shorter distance (2.01 Å) leads to an increased interaction between the CO$_2$ and the cobalt, with the latter being pulled up and becoming visible as a bright dot in the nc-AFM simulation (**b**). Larger distance (3.37 Å) produces a nc-AFM image with the linear feature due to the CO$_2$ that is too far from the tpy core as compared to the experimentally observed configuration. All these structures are energetically very similar, with their total energy values falling within 0.017 eV. The fact that the cobalt is not significantly pulled up in the case that better reproduces the experimental image confirms that the interaction between such atom and the CO$_2$ molecule is rather weak (tpy-Co···CO$_2$). Scale bars: 0.5 nm.

**CO$_2$ conversion into CO**

In the captivating encounter between our SAC and CO$_2$, not only did we witness the presence of individual gas molecules positioned on the surface and in proximity to active sites but also observed them diminishing in favor of CO. This revelation suggests the intriguing possibility of CO$_2$ transforming into CO. Such a conversion, although unexpected, is deduced from our compelling results and meticulously conducted control experiments, as described below.

A sample exposed to CO$_2$ while being held at 9.5 K reveals 54% of tpy-Co···CO$_2$ sites along with 9% of tpy-Co-CO ones. Successive WA steps (see Methods) at increasingly longer times promoted changes in the active site composition (Supplementary Fig. 9 and Supplementary Table 3). Surprisingly, the relative abundance of tpy-Co-CO sites steadily increased, reaching a maximum of 63% after 7 minutes of WA and replacing the "empty" tpy-Co sites (Supplementary Fig. 9, WA of 3 to 7 minutes). In 11 minutes, the number of tpy-Co-CO sites slightly decreased, and dropped to 8% after 30 minutes, as the sample likely reached a temperature high enough to promote a significant CO desorption. Remarkably, a further WA of 3 minutes (last point in Supplementary Fig. 9) does not lead to any increase of tpy-Co-CO sites.

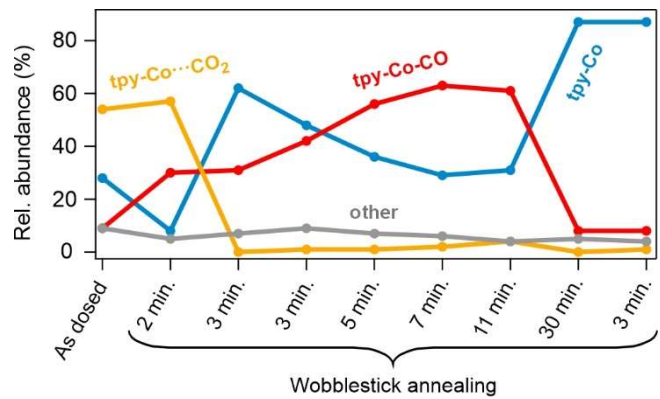

**Supplementary Fig. 9 | Evolution of the active sites' occupation.** Relative abundance of the tpy-Co sites with different occupation as a function of the residence time of the sample on the wobblestick in consecutive WA steps (see Methods). Data are plotted from the values in Supplementary Table 3.

**Supplementary Table 3 | Statistical analysis of the active site composition in $CO_2$ experiments.**

| Active site composition | As dosed | WA 2 min. | WA 3 min. | WA 3 min. | WA 5 min. | WA 7 min. | WA 11 min. | WA 30 min. | WA 3 min. |
|---|---|---|---|---|---|---|---|---|---|
| tpy-Co⋯$CO_2$ | 54% | 57% | 0% | 1% | 1% | 2% | 4% | 0% | 1% |
| tpy-Co-CO | 9% | 30% | 31% | 42% | 56% | 63% | 61% | 8% | 8% |
| tpy-Co | 28% | 8% | 62% | 48% | 36% | 29% | 31% | 87% | 87% |
| others | 9% | 5% | 7% | 9% | 7% | 6% | 4% | 5% | 4% |

Relative abundance of the main species as a function of the sample preparation and treatment. The initial sample was obtained by exposing the SAP to $CO_2$ at $5\times10^{-8}$ mbar for 15 minutes on the SAP-decorated sample at 9.5 K. Subsequent wobblestick annealing (WA) steps were performed, holding the sample on the wobblestick pincer for increasingly longer times. Some features at the active sites could not be assigned clearly and were categorized as "others". The data in this table were obtained by counting between 500 and 1000 active sites at each step and were used to produce the graph in Supplementary Fig. 9.

We conducted a comprehensive set of targeted control experiments (see below) to exclude other possible sources of CO and can speculate that a $CO_2$ conversion is a likely option. With reference to Supplementary Fig. 9, the disappearance of tpy-Co⋯$CO_2$ sites after 3 minutes of WA in favor of the "empty" tpy-Co ones seems to discard a possible $CO_2$ conversion to CO, as the latter (namely tpy-Co-CO sites) keeps increasing also when there are no longer tpy-Co⋯$CO_2$ moieties (Supplementary Fig. 9, WA of 3 to 7 minutes). However, we point out that our statistical analysis only considers the species bound (or sitting next) to the tpy-Co sites and does not take into account those present in other parts of the surface. These are difficult to identify but could represent a reservoir of molecules that diffuse around and interact with the tpy-Co sites during the annealing steps. In this perspective, one could hypothesize the following scenario: during the WA steps, part of the $CO_2$ molecules that initially sat next to the polymer chains is converted to CO and bind to the cobalt atoms, while others diffuse away and get stabilized somewhere else on the surface (e.g. at step edges and defects). At each consecutive WA, some of the $CO_2$ molecules diffuse away from these other locations, occasionally encounter the available tpy-Co sites, and eventually convert into CO that, again, remains bound to the cobalt. As we have discussed in the main text, CO binds to the cobalt atoms stronger than the $CO_2$, which is only weakly stabilized next to it. This is in line with the fact that $CO_2$ molecules have the freedom to diffuse away from the cobalt atoms, while CO tends to strongly bind it. After a long WA of 30 minutes all CO and $CO_2$ likely desorb from the surface and even an additional WA of 3 minutes does not lead to any increase of tpy-Co-CO sites. We highlight that the initial absolute amount of $CO_2$ on the surface is larger than the number of tpy-Co sites, as displayed by the representative STM image in Supplementary Fig. 10. This observation could support the hypothesis of having a reservoir of $CO_2$ in other locations of the surface, and readily available during the annealing steps.

Finally, it must be noted that the typical rest gas in UHV setups presents abundant $H_2$ amounts. Molecular hydrogen is a known reductant for $CO_2$. The reaction $CO_2 + H_2 \rightarrow CO + H_2O$ has an enthalpy of 41 kJ·mol$^{-1}$, much lower than the 532 kJ·mol$^{-1}$ of the $CO_2$ dissociation into CO and O, thus suggesting a plausible scenario for the observed chemical transformations[S2].

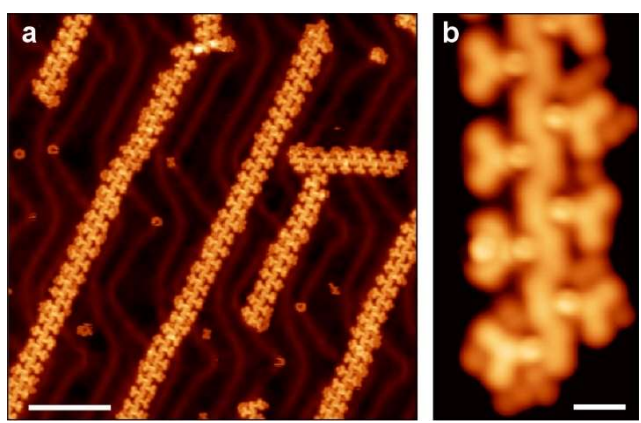

**Supplementary Fig. 10 | $CO_2$-dosed sample.** STM images after $CO_2$ exposure at 5×10$^{-8}$ mbar for 15 minutes on the SAP-decorated sample at 9.5 K. The spherical protrusions that are stabilized next to the polymer chains are $CO_2$ molecules. Despite an excess of such features with respect to the active sites of the polymers, some tpy-Co sites appear "empty". The circular features on the bare gold surface are clusters of self-assembled $CO_2$ molecules. A wobblestick annealing of 2 minutes is necessary to promote a significant increase of the tpy-Co⋯$CO_2$ sites (see Supplementary Fig. 1f) and the disappearance of the excess of $CO_2$ from the locations next to the polymer chains. Scanning parameters: $I_t$ = 100 pA, $V_b$ = –0.10 V (**a**); $I_t$ = 70 pA, $V_b$ = –0.20 V (**b**). Scale bars: 10 nm (**a**), 1 nm (**b**).

**Control experiments**

The observation of tpy-Co-CO sites after annealing a $CO_2$-dosed SAP is remarkable, and it could be the signature of a $CO_2$ reduction to CO. However, the experimental vacuum chamber used for the reported experiments is constantly dosed with CO for tip functionalization purposes, and there might be different origin of the CO manifested as tpy-Co-CO moieties. Therefore, we performed several control experiments to exclude other possible sources of CO, as outlined below.

1. *Blank*
   A sample featuring tpy-Co sites was inserted in the cold STM stage, let cool down, and then the same processing as in real $CO_2$ exposures was performed, but without introducing any $CO_2$ gas in the chamber. Specifically, when the sample reached the base temperature of 4.7 K, we opened the cryostat shields, waited 15 minutes, and then closed the shields. STM investigation of the resulting sample revealed the complete absence of tpy-Co-CO and tpy-Co⋯$CO_2$ sites. This result should be compared to the "As dosed" sample in Supplementary Table 1, where 54% of the tpy-Co sites were occupied by $CO_2$ and 9% were in the form of tpy-Co-CO, demonstrating that in absence of any $CO_2$ inlet there is correspondingly no CO.

2. *Wobblestick annealing of 3 minutes*
   The sample obtained from the previous treatment (blank experiment) was extracted from the cold STM stage, let warm up on the wobblestick for 3 minutes, and inserted again in the STM stage. STM investigation of the resulting sample revealed 1% of tpy-Co-CO and 5% of tpy-Co···$CO_2$ sites. These values are within the experimental error of the yield determination analysis, and are considered as negligible. Moreover, some other impurities show sometimes a similar appearance of the tpy-Co···$CO_2$ sites, and could be confused, justifying the overestimated value of 5%. In particular, the same sample treatment after dosing $CO_2$ on the sample led to 31% of tpy-Co-CO sites as compared to the 1% of the present case, which clearly highlights how the tpy-Co-CO sites increased only if $CO_2$ was previously dosed.

3. *Argon dosing + wobblestick annealing of 11 minutes*

   Here, we excluded that some CO molecules could reach the surface of the sample after being displaced from the cryostat walls and STM stage due to $CO_2$ collisions with these surfaces. To do so, we dosed another gas, namely argon, with the same exposure used during the $CO_2$ experiments. After such Ar dosing on a standard tpy-Co sample, we performed a WA for 11 minutes, which is expected to maximize the abundance of tpy-Co-CO sites (61% in the experiment with $CO_2$, see Supplementary Table 1). However, we only found 4% of tpy-Co-CO sites.

4. *Ion gauge off*

   In some cases, ionization processes occurring in the proximity of ion gauge filaments (used to monitor the pressure in UHV systems) can induce chemical reactions in the molecules composing the rest gas of the vacuum setup. Here, we excluded that the CO was produced by the ionization of $CO_2$ from such filament, repeating the experiment with the ion gauge switched off. After cooling down the sample with tpy-Co sites in the STM stage we exposed it to $CO_2$ as in previous experiments, but keeping the ion gauge off. After the exposure, we performed a wobblestick annealing of 11 minutes to promote the increase of tpy-Co-CO sites. As expected, these sites amounted to 83%, demonstrating that the CO does not originate from $CO_2$ cracking operated by the ion gauge filament. The significantly higher value as compared to the 61% reported in Supplementary Table 1 could be due to a slightly higher $CO_2$ partial pressure during the exposure, which could not be kept entirely constant in absence of any pressure readout.

**NMR spectra**

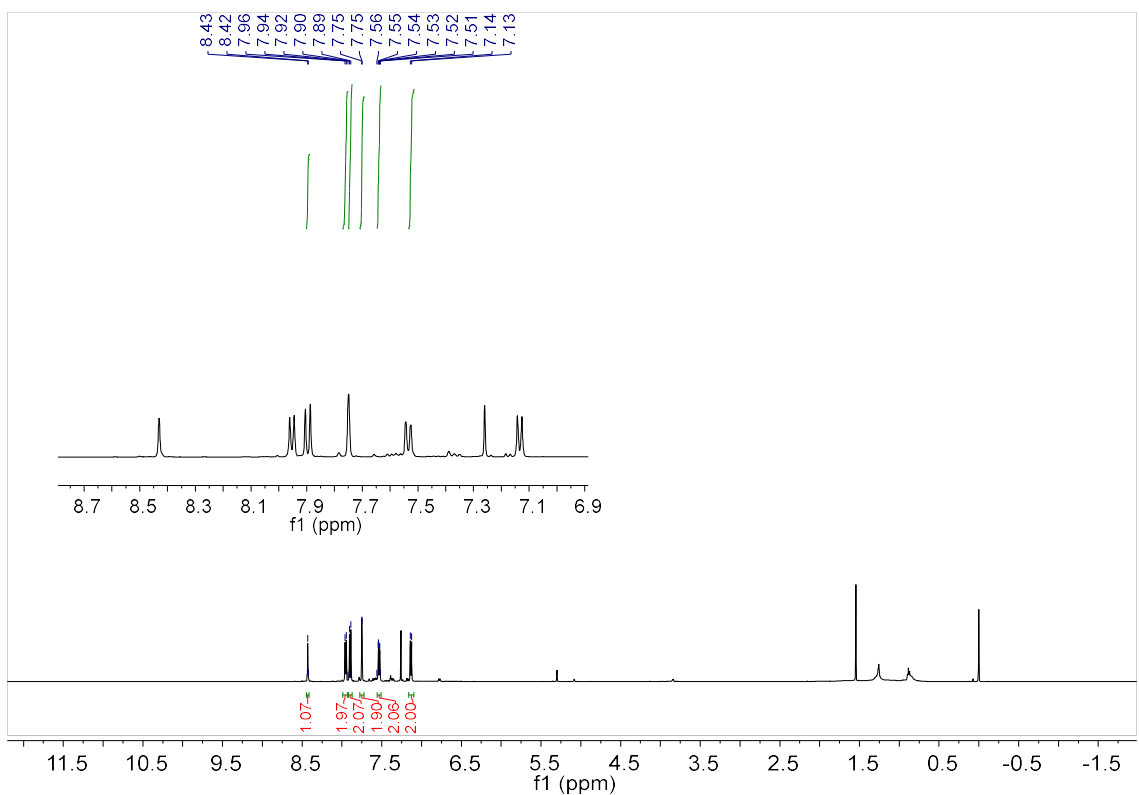

**Supplementary Fig. 11** | $^1$H NMR spectrum of compound **4** in CDCl$_3$ (500 MHz, 298 K).

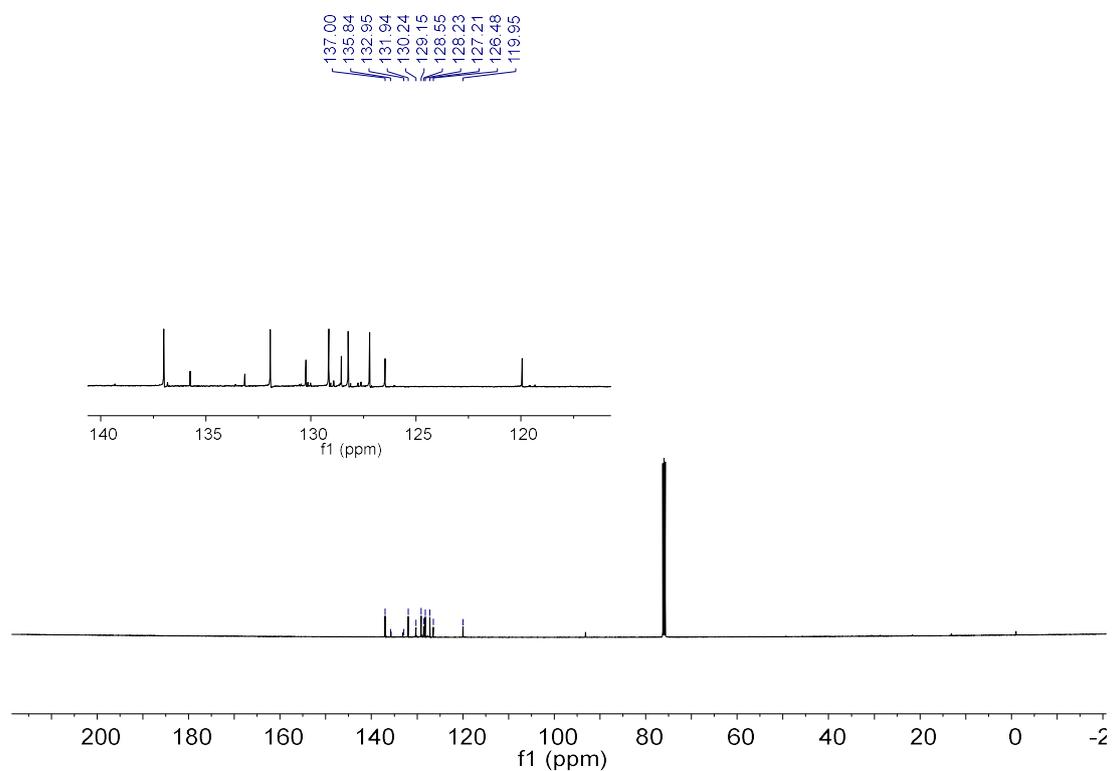

**Supplementary Fig. 12** | $^{13}$C NMR spectrum of compound **4** in CDCl$_3$ (126 MHz, 298 K).

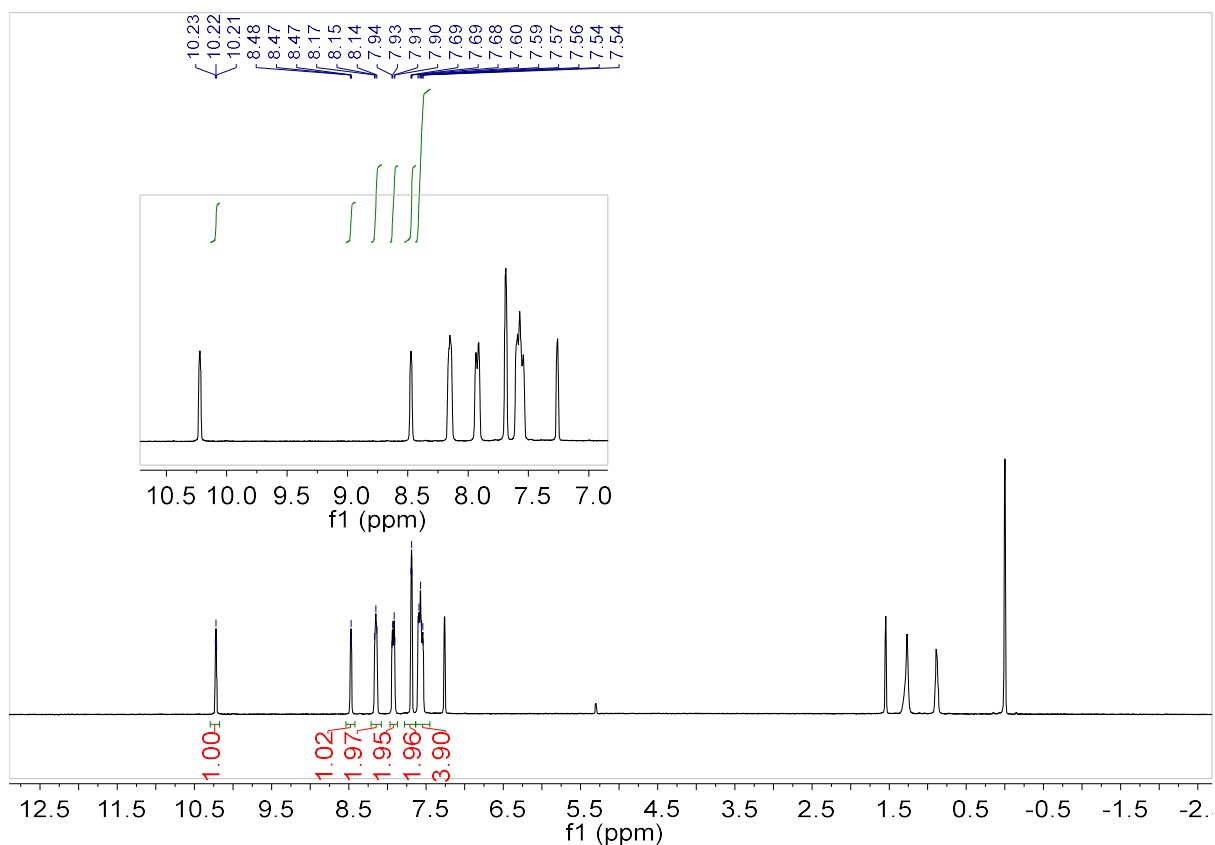

**Supplementary Fig. 13** | $^1$H NMR spectrum of compound **5** in CDCl$_3$ (400 MHz, 298 K).

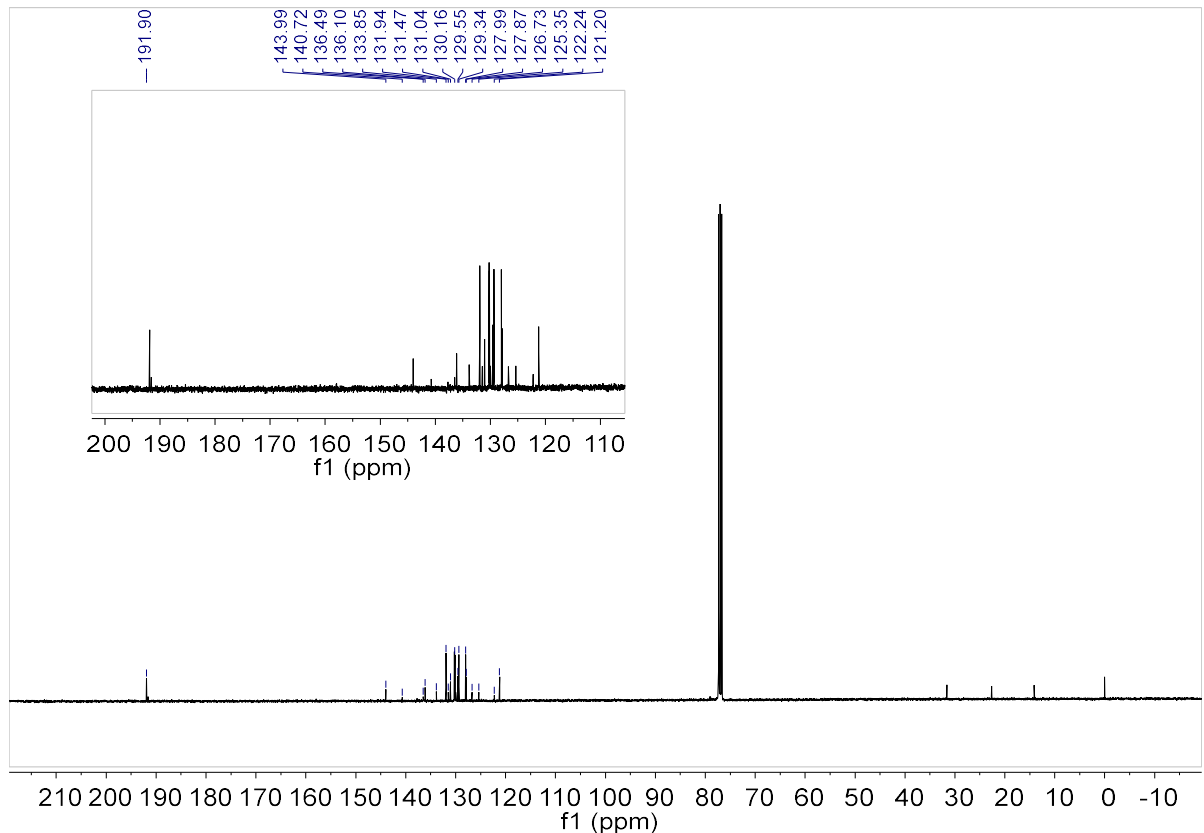

**Supplementary Fig. 14** | $^{13}$C NMR spectrum of compound **5** in CDCl$_3$ (101 MHz, 298 K).

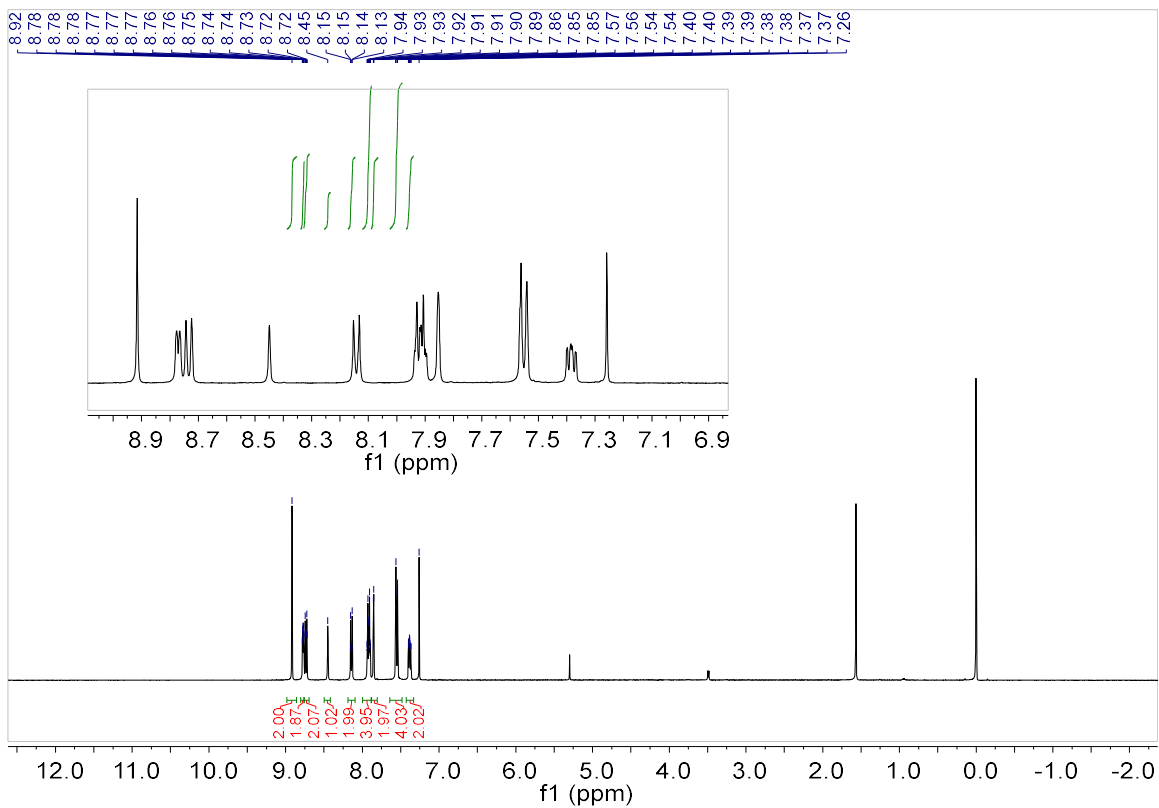

**Supplementary Fig. 15** | ¹H NMR spectrum of compound **1** in CDCl$_3$ (400 MHz, 298 K).

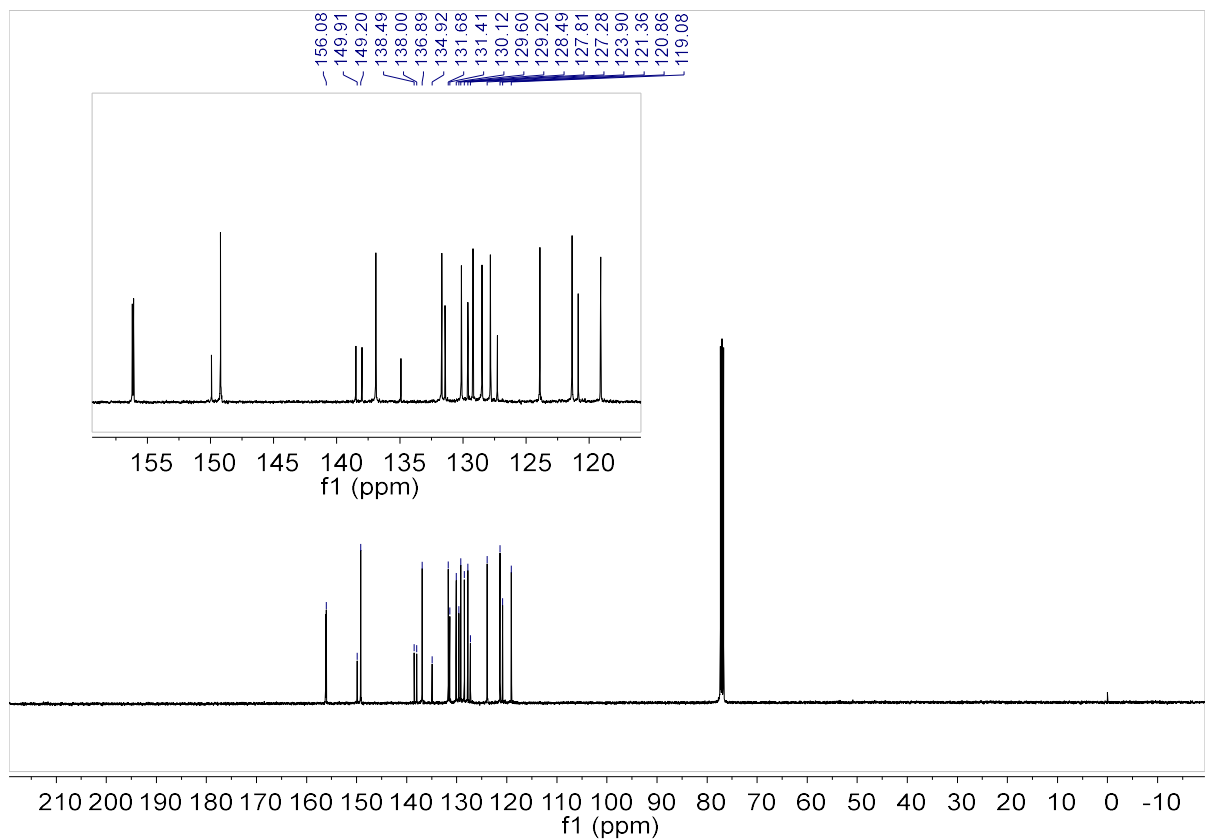

**Supplementary Fig. 16** | ¹³C NMR spectrum of compound **1** in CDCl$_3$ (101 MHz, 298 K).

**Supplementary references**